\documentstyle[12pt]{article}
\begin{document}
\newcommand{\siml}{\stackrel{<}{\sim}}
\newcommand{\simg}{\stackrel{>}{\sim}}
\newcommand{\lleq}{\stackrel{<}{=}}

\baselineskip=1.333\baselineskip


%
\begin{center}
{\large\bf
Synchronizations 
in small-world networks of spiking neurons:
Diffusive versus sigmoid couplings
} 
\end{center}

\begin{center}
Hideo Hasegawa
\footnote{e-mail:  hasegawa@u-gakugei.ac.jp}
\end{center}

\begin{center}
{\it Department of Physics, Tokyo Gakugei University  \\
Koganei, Tokyo 184-8501, Japan}
\end{center}
\begin{center}
({\today})
\end{center}
\thispagestyle{myheadings}

\begin{abstract}
By using a semi-analytical dynamical mean-field
approximation previously proposed by the author
[H. Hasegawa, Phys. Rev. E ,
{\bf 70}, 066107 (2004)],
we have studied the synchronization of 
stochastic, small-world (SW) networks of FitzHugh-Nagumo
neurons with diffusive couplings.
The difference and similarity between
results for {\it diffusive} and {\it sigmoid} couplings have 
been discussed.
It has been shown that with introducing
the weak heterogeneity to regular networks,
the synchronization may be slightly increased
for diffusive couplings,
while it is decreased for sigmoid couplings.
This increase in the synchronization for diffusive couplings
is shown to be due to their local, negative feedback contributions, 
but not due to the shorten average distance in SW networks. 
Synchronization of SW networks depends
not only on their structure but also
on the type of couplings.

\end{abstract}

\vspace{0.5cm}

%
\newpage
\section{INTRODUCTION}


In recent years, much attention has been paid to
complex networks such as small-world (SW) 
and scale-free (SF) networks
(for a review see \cite{Alb02,Dor02}). 
The SW network, which has been 
proposed by Wattz and Strogatz (WS) \cite{Wat98,Str01},
is characterized by a large clustering 
coefficient and a small average distance.
The original WS-SW network has been created by
introducing the finite-degree heterogeneity
to the regular network by random rewirings of links. 
Newman and Wattz (NW) \cite{New99} have proposed an alternative
SW network, randomly adding shortcut links
to the regular networks without rewirings. 
In SW networks, the degree distribution $P(k)$ for a node to
have $k$ coupled neighbors has
an exponential tail for a large $k$.
In contrast, the degree distribution in SF networks, which was
first proposed by  
Barab\'{a}si and Albert \cite{Bara99},
is given by the power law as 
$P(k) \sim k^{-\gamma}$ with the index $\gamma$ ($= 2 \sim 4$).
Since Barab\'{a}si and Albert \cite{Bara99} have proposed
a growing SF network with preferential
attachments of nodes,
many models and mechanisms have been proposed 
not only for growing but also for non-growing SF networks 
with geographical and non-geographical structures.

The interplay between structure and dynamics has attracted
a great deal of attention
to the synchronization in complex networks.
The synchronization in SW networks 
consisting of spiking neurons has been studied 
\cite{Fernandez00}-\cite{Buz04}
with the use of Hodgkin-Huxley (HH) 
\cite{Fernandez00}\cite{Net04}, 
FitzHugh-Nagumo (FN) \cite{Fernandez01}\cite{Hasegawa04}
Hindmarsh-Rose (HR) \cite{Bucolo02}, 
integrate-and-fire (IF) \cite{Net04}\cite{Mas04},
and phase models \cite{Hon02a,Hon02b}. 
By using a more general class of models,
dynamical properties including
synchrony in SW and SF networks 
have been also investigated
\cite{Barahona02,Nishikawa03,Mot05}.
It has been, however, controversial whether the synchronization
in complex networks is better or worse than that in regular networks. 
Most of calculations have shown that
the synchronization in SW networks is better
than regular networks 
because of the shorten average distance in the former
\cite{Fernandez00,Bucolo02,Hon02a,Hon02b,Buz04,Barahona02}.
On the contrary, it has been shown 
that the average distance is not necessarily correlated
with the synchronizability of the networks
\cite{Hasegawa04}\cite{Nishikawa03}.
Some have claimed that the synchronization is increased or
decreased depending on the adopted parameters or
calculation conditions \cite{Net04}\cite{Fernandez01}\cite{Mas04}.

In a previous paper \cite{Hasegawa04},
we have developed a semi-analytical theory
for SW networks,
by generalizing the dynamical mean-field approximation (DMA)
which was originally proposed for regular networks
with all-to-all couplings \cite{Hasegawa03a,Hasegawa03b}.
The method newly developed in \cite{Hasegawa04} is applicable 
to SW networks with a wide range of couplings
covering from local to global and/or from
regular to random ones.
In \cite{Hasegawa04}, we have taken into account
three kinds of spatial correlations: on-site correlation,
the correlation for a coupled pair, and
that for a pair without direct couplings.
Our method has been applied  
to SW FN neural networks with sigmoid couplings
\cite{Hasegawa04},
in which a coupling to the neuron $i$ is given by
\begin{equation}
I_i^{(c)}(t) = J \sum_j c_{ij}\: G(x_j(t)),
\end{equation}
where $J$ denotes the coupling strength,
$G(x)=1/[1+exp(-(x-\theta)/\alpha)]$ 
is the sigmoid function with threshold $\theta$
and width $\alpha$, and
the adjacent matrix $c_{ij}$ is $c_{ij}=c_{ji}=1$ for 
a coupled $(i,\:j)$ pair and 0 otherwise.
Calculations by DMA 
and direct simulations
have shown that when random links are
added to regular networks,
the synchronization is decreased because
of the introduced heterogeneity in SW networks
\cite{Hasegawa04}.

Besides the sigmoid coupling of Eq. (1),
the diffusive coupling given by
\begin{equation}
I_i^{(c)}(t) = K \sum_j c_{ij}\: H(x_j(t)-x_i(t)),
\end{equation}
has been widely employed for theoretical study
on neural networks,
where $K$ stands for the coupling strength
and $H(x-y)$ the coupling function.
Equations (1) and (2) model chemical and electrical
synapses, respectively.
Both chemical and electrical synapses exist
in neocortex.
Chemical synapses use a chemical neurotransmitter
that is packaged presynaptically into vesicle,
released in quantized amount, and
binds to postsynaptic receptors.
In contrast, electrical synapses are simpler
in structure and function. They provide
a direct pathway that allows ionic current to
flow from the cytoplasm of one cell to that of another \cite{Gib03}.
Although chemical synapses are by far the most abundant,
electrical synapses also play an important role in neocortex.
The purpose of the present paper is to 
apply the DMA to SW neural networks of FN neurons 
with diffusive couplings,
and to compare the
results for diffusive couplings to those
for sigmoid couplings in \cite{Hasegawa04}.
This is expected to provide some insight to unsettled
issue on the effect of the heterogeneity on the
synchronization in SW networks mentioned above.

The paper is organized as follows.
In Sec. II, we have derived differential equations (DEs), 
applying the DMA
to SW FN networks with
diffusive couplings in order to transform 
the original stochastic DEs to deterministic DEs.
In Sec. 3.1 and 3.2, we have reported numerical calculations
for regular and SW networks, respectively.
The final Sec. IV is devoted to conclusion and discussion. 

\section{Small-world networks of FN neurons}

\subsection{Adopted model and method}

We have assumed that $N$-unit FN neurons
are distributed on a ring 
with the average coordination number $Z$ and 
the coupling randomness $p$.
Dynamics of a single neuron $i$ in a given SW network
is described by the non-linear DEs given by 
\begin{eqnarray}
\frac{dx_{1i}(t)}{dt} &=& F[x_{1i}(t)]
- c \:x_{2i}(t)
+I_i^{(c)}(t)
+I_{i}^{(e)}(t)+\xi_i(t), \\
\frac{dx_{2i}(t)}{dt} &=& b \:x_{1i}(t) - d \:x_{2i}(t)+e,
\hspace{2cm}\mbox{($i=1$ to $N$)}
\end{eqnarray}
with
\begin{eqnarray}
I_i^{(c)}(t)&=& K \sum_j\: c_{ij}\:H(x_{1j}(t)-x_{1i}(t)), \\ 
I_i^{(e)}(t)&=& A \:\Theta(t-t_{in})\:\Theta(t_{in}+t_{w}-t).
\end{eqnarray}
In Eq. (3)-(6),
$F[x(t)]=k\: x(t)\: [x(t)-a]\: [1-x(t)]$, 
$k=0.5$, $a=0.1$, $b=0.015$, $d=0.003$ and $e=0$
\cite{Hasegawa03a}\cite{Rod96}\cite{Tuckwell98}: 
$x_{1i}$ and $x_{2i}$ denote the fast (voltage) variable
and slow (recovery) variable, respectively:
$H(x)$ stands for the diffusive-type coupling:
$c_{ij}$ the adjacent matrix given by $c_{ij}=c_{ji}=1$ for
a coupled $(i,\:j)$ pair and zero otherwise,
self-coupling terms being excluded ($c_{ii}=0$).
By changing $Z$ value, 
our model given by Eqs. (3)-(6) covers from local couplings ($Z \ll N$)
to global couplings ($Z=N-1$).
We should, however, keep in mind that the electrical synapses
by nature can only be produced among close neurons.
The response of neuron networks has been studied to
an external, single spike input given by $I_i^{(e)}(t)$
with magnitude $A$ and spike width $t_w$ applied for 
$t_{in} \leq t < t_{in}+t_{w}$, $\Theta(x)$ being
the Heaviside function.
Added white noises $\xi_i(t)$ are given by
\begin{eqnarray}
<\xi_i(t)>&=&0, \\ 
<\xi_i(t)\:\xi_j(t')>&=&\beta^2 \; \delta_{ij}\:\delta(t-t'),
\end{eqnarray}
where the average of $<U({\bf z},t)>$
for an arbitrary function of $U({\bf z},t)$ is given by
\begin{equation}
<U({\bf z},t)>
= \int...\int \;d{\bf z} \;U({\bf z},t) \;Pr({\bf z}),
\end{equation}
$Pr({\bf z})$ denoting a probability distribution function 
for $2N$-dimensional
random variables ${\bf z}=(\{ x_{\kappa i} \})$.

Our WS-SW network has been made after \cite{Wat98}.
Starting from a regular network, 
$N_{ch}$ couplings among $N Z/2$ couplings
are randomly modified such that $c_{ij}=1$
is changed to $c_{ij}=0$ or vice versa.
The coupling randomness $p$ is given by
\begin{equation}
p = \frac{2N_{ch}}{N Z},
\end{equation}
which is 0 and 1 for completely regular and random couplings,
respectively.

In DMA \cite{Hasegawa03a}\cite{Hasegawa04}, 
we will obtain equations of motions for means,
variances and covariances of state variables.
Variables spatially averaged over the ensemble are defined by
\begin{eqnarray}
X_{\kappa}(t)&=&\frac{1}{N}\;\sum_{i} \;x_{\kappa i},
\hspace{1cm}\mbox{$\kappa=1,\:2$}
\end{eqnarray}
and their means by
\begin{eqnarray}
\mu_{\kappa}(t)&=&\left< \left< 
X_{\kappa}(t) \right> \right>_c,
\end{eqnarray}
where the bracket $< \cdot >_c$ 
denotes the average over the coupling configuration.
As for variances and covariances of state variables,
we consider three kinds of spatial correlations:
(1) on-site correlation ($\gamma$),
(2) the correlation for a coupled pair ($\zeta$),
and (3) that for a pair without direct couplings ($\eta$):
\begin{equation}
\left< \left< \delta x_{\kappa i}
\: \delta x _{\lambda j}\right> \right>_c = \left\{
  \begin{array}{ll}
    \gamma_{\kappa,\lambda}, &\mbox{for $i=j$}\\
    \zeta_{\kappa,\lambda}, &\mbox{for $i\neq j,\; c_{ij}=1$}\\
    \eta_{\kappa,\lambda}, &\mbox{for $i \neq j,\; c_{ij}=0$}
  \end{array}\right.
\end{equation} 
where $\kappa, \lambda=1,\:2$ and
\begin{eqnarray}
\delta x_{\kappa i}(t)&=& x_{\kappa i}(t)-\mu_{\kappa}(t).
\end{eqnarray}
In Eq. (13), $\gamma_{\kappa,\lambda}$, 
$\zeta_{\kappa,\lambda}$ and 
$\eta_{\kappa,\lambda}$ are defined by
\begin{eqnarray}
\gamma_{\kappa,\lambda}(t)
&=&\left< \frac{1}{N} \sum_i \left<\delta x_{\kappa i}(t)
\:\delta x_{\lambda i}(t)\: \right> \right>_c, \\
\zeta_{\kappa,\lambda}(t)
&=& \left< \frac{1}{N Z} \sum_i \;\sum_{j}
c_{ij} \left< \delta x_{\kappa i}(t)
\;\delta x_{\lambda j}(t) \right> \right>_c, \\
\eta_{\kappa,\lambda}(t)
&=& \left< \frac{1}{N (N-Z-1)} \sum_i \;\sum_{j}
(1-\delta_{ij} - c_{ij}) 
\left< \delta x_{\kappa i}(t)
\;\delta x_{\lambda j}(t) \right> \right>_c.
\end{eqnarray}
For a later purpose, we define also the spatially-averaged
correlation given by
\begin{eqnarray}
\rho_{\kappa,\lambda}(t)&=& \left< \frac{1}{N^2} \sum_i \;\sum_{j}
\left< \delta x_{\kappa i}(t)
\;\delta x_{\lambda j}(t) \right> \right>_c,\\
&=& \left< \left<\delta X_{\kappa}(t) 
\:\delta X_{\lambda}(t) \right> \right>_c,
\end{eqnarray} 
where $\delta X_{\kappa}(t)=X_{\kappa}(t)-\mu_{\kappa}(t)$.
It is noted that $\gamma_{\kappa,\lambda}$, 
$\zeta_{\kappa,\lambda}$
$\eta_{\kappa,\lambda}$ and 
$\rho_{\kappa,\lambda}$ are not independent,
obeying the sum rule given by
\begin{equation}
N \rho_{\kappa,\lambda}
= \gamma_{\kappa,\lambda}+ Z \zeta_{\kappa,\lambda}
+ (N-Z-1) \eta_{\kappa,\lambda}.
\end{equation}
In order to derive Eqs. (15)-(20), 
we have employed the decomposition:
\begin{eqnarray} 
1&=&\delta_{ij}+(1-\delta_{ij})[c_{ij}+(1-c_{ij})], \nonumber \\
&=& \delta_{ij}+c_{ij}+(1-\delta_{ij}-c_{ij}), 
\end{eqnarray}
with $c_{ii}=0$.

In calculating means, variances and covariances given 
by Eqs. (12) and (15)-(18),
we have assumed that (1) the noise intensity is weak, and
(2) the distribution of state variables takes the
Gaussian form.
By using the first assumption, we expand DEs given by Eqs. (3)-(6)
in a power series of fluctuations around means.
The second assumption may be justified by some numerical calculations
for stochastic FN \cite{Tuckwell98}\cite{Tanabe01} and 
HH neuron models \cite{Tanabe99}\cite{Tanabe01a}. 
It has been shown that for weak noises, the distribution
of a membrane potential of a single FN or HH
neuron nearly follows the Gaussian distribution,
although for strong noises, the distribution deviates
from the Gaussian, taking a bimodal form.

Before closing Sec. 2.1, we briefly summarize the
introduced variables and their meanings as follows:
$N$, the number of neurons:
$Z$, the average coordination number:
$p$, the coupling randomness:
$K$, the coupling strength:
$c_{ij}$, the adjacent matrix:
$X_{\kappa}$, the spatially average of the fast ($\kappa=1$)
and slow ($\kappa=2$) variables;
$\mu_{\kappa}$, a mean value of $X_{\kappa}$;
$\gamma_{\kappa,\lambda}$,
$\zeta_{\kappa,\lambda}$, and
$\eta_{\kappa,\lambda}$, 
the correlations of on-site, a coupled pair,
and an uncoupled pair, respectively.

\subsection{Equations of motions}

We will obtain equations of motions for 
$\mu_{\kappa}(t)$, $\gamma_{\kappa,\lambda}(t)$, 
$\zeta_{\kappa,\lambda}(t)$, $\eta_{\kappa,\lambda}(t)$
and $\rho_{\kappa,\lambda}(t)$. 
Readers who are not interested in mathematical details, 
may skip to Sec. 2.3, where our theoretical results
are summarized.

After some manipulations, we get the following DEs
(the  argument $t$ being suppressed;
for details, see appendix A):
\begin{eqnarray}
\frac{d \mu_1}{d t}&=&f_0 + f_2 \gamma_{1,1} -c \mu_2
+I_{ext}, \\
\frac{d \mu_2}{d t}&=& b \mu_1 - d \mu_2 +e,  \\
\frac{d \gamma_{1,1}}{d t}&=& 2 (a \gamma_{1,1}- c \gamma_{1,2})
+ 2 K Z h_1 (\zeta_{1,1}-\gamma_{1,1})
+\beta^2, \\
\frac{d \gamma_{2,2}}{d t}&=& 2 (b \gamma_{1,2}- d \gamma_{2,2}),  \\
\frac{d \gamma_{1,2}}{d t}&=& b \gamma_{1,1}
+ (a-d) \gamma_{1,2} 
- c \gamma_{2,2}
+ K Z h_1 (\zeta_{1,2}-\gamma_{1,1}),  \\
\frac{d \rho_{1,1}}{d t}&=& 2 (a \rho_{1,1} - c \rho_{1,2})
+\frac{\beta^2}{N}, \\
\frac{d \rho_{2,2}}{d t}&=& 2 (b \rho_{1,2}- d \rho_{2,2}), \\
\frac{d \rho_{1,2}}{d t}&=& b \rho_{1,1}
+ (a-d) \rho_{1,2} - c \rho_{2,2}, \\
\frac{d \zeta_{1,1}}{d t}&=& 2 (a \zeta_{1,1} - c \zeta_{1,2}) \nonumber\\
&+&  2 K h_1 [\gamma_{1,1}+(ZC-ZR)\zeta_{1,1}
+(ZR-ZC-1)\eta_{1,1}],\\
\frac{d \zeta_{2,2}}{d t}&=& 2 (b \zeta_{1,2}- d \zeta_{2,2}),  \\
\frac{d \zeta_{1,2}}{d t}&=& b \zeta_{1,1}
+ (a-d) \zeta_{1,2} - c \zeta_{2,2} \nonumber\\
&+& K h_1 [\gamma_{1,2}+(ZC-ZR)\zeta_{1,2}
+(ZR-ZC-1)\eta_{1,2}],\\
\frac{d \eta_{1,1}}{d t}&=& 2 (a \eta_{1,1} - c \eta_{1,2})\nonumber \\
&+&\left(\frac{2 K Z h_1}{N-Z-1}\right) 
[(ZR-ZC-1) (\zeta_{1,1}-\eta_{1,1})], \\
\frac{d \eta_{2,2}}{d t}&=& 2 (b \eta_{1,2}- d \eta_{2,2}),  \\
\frac{d \eta_{1,2}}{d t}&=& b \eta_{1,1}
+ (a-d) \eta_{1,2} - c \eta_{2,2} \nonumber \\
&+& \left(\frac{K Z h_1}{N-Z-1} \right) 
[(ZR-ZC-1)(\zeta_{1,2}-\eta_{1,2})],
\end{eqnarray}
with
\begin{eqnarray}
C&=& \left< \frac{1}{NZ^2} 
\sum_i \sum_j \sum_k \:c_{ij}\: c_{jk}\: c_{ik} \right>_c,\\
R&=& \left< \frac{1}{N Z^2} 
\sum_i \sum_j \sum_k c_{ij}\: c_{jk} \right>_c, 
\end{eqnarray}
where $f_{\ell}=(1/ /\ell ! ) \:F^{(\ell)}$ and $h_1 = H^{(1)}(0)$
with $H(0)=H^{(2)}(0)=0$.

\subsection{Summary of our method}

The clustering coefficient $C$ and
the coupling connectivity $R$, which are
given by Eqs. (36) and (37), respectively, 
play important roles in our DMA theory for SW networks. 
The clustering coefficient $C$
introduced in SW networks \cite{Wat98,Str01},
expresses a factor forming a cluster 
where the three sites $i$, $j$ and $k$
are mutually coupled.
In contrast, the coupling connectivity $R$
expresses a factor for a cluster where the two sites $j$ and $k$
are coupled to the third site $i$, but the sites $j$ and $k$ 
are not necessarily coupled \cite{Note1}.
It is noted in Eqs. (22)-(35)
that there is no explicit dependence on 
the coupling randomness $p$,
and it is only through parameters $C$ and $R$
in the mean-field equations.

Figures 1(a) and 1(b) show the $p$ dependences of
$C$ and $R$, respectively, for various $Z$ values
with $N=100$.
With increasing $p$ from zero,
$C$ is decreased and approaches $C = Z/N$ at $p = 1$.
In contrast, $R$ is monotonically increased with increasing $p$.

Among the 12 correlations such as $\gamma_{\kappa,\lambda}$ {\it et al.}
given by Eqs. (15)-(18), 
nine correlations are independent because of the
sum rule given by Eq. (20). In this study,
we have chosen nine correlations of $\gamma_{\kappa,\lambda}$,
$\zeta_{\kappa,\lambda}$ and $\rho_{\kappa,\lambda}$
as independent variables.
Then the original $2 N$-dimensional {\it stochastic}
DEs given by Eqs.(3) and (4) have been transformed to
11-dimensional {\it deterministic} DEs.
Equations of motions for diffusive couplings given by
Eqs. (22)-(35) are rather different from those
for sigmoid couplings given by Eqs. (21)-(34) 
in \cite{Hasegawa04}, 
related discussion being given in Sec. 4.


From a comparison of Eq. (24) with Eq. (27),
we note that
\begin{eqnarray}
\rho_{1,1} &=& \frac{\gamma_{1,1}}{N},
\hspace{1cm}\mbox{for $K/\beta \rightarrow 0$} \\
&=& \gamma_{1,1},
\hspace{1cm}\mbox{for $\beta/K \rightarrow 0$}
\end{eqnarray}
where Eq. (38) is nothing but the central-limit theorem
describing the relation between fluctuations
of local and average variables [Eqs. (15) and (19)].
In order to quantitatively discuss the
synchronization, we first consider the quantity given by
\begin{equation}
P(t)=\frac{1}{N^2} \sum_{i j}<[x_{1i}(t)-x_{1j}(t)]^2>
=2 [\gamma_{1,1}(t)-\rho_{1,1}(t)].
\end{equation}
When all neurons are in the completely synchronous state,
we get $x_{1i}(t)=X_1(t)$ for all $i$, and 
then $P(t)=0$ in Eq. (40).
On the contrary, in the asynchronous state, we get 
$P(t)=2(1-1/N)\gamma_{1,1} \equiv P_0(t)$
from Eq. (38).
We have defined the synchronization ratio
given by \cite{Hasegawa04}
\begin{equation}
S(t) =1-\frac{P(t)}{P_0(t)}
= \left( \frac{N\rho_{1,1}(t)/\gamma_{1,1}(t)-1}{N-1} \right),
\end{equation}
which is 0 and 1 for completely asynchronous ($P=P_0$)  
and synchronous states ($P=0$), respectively.

We define the time $t_{max}$
when $S(t)$ takes its maximum value as
\begin{equation}
t_{max}=\{t \mid d S(t)/d t=0, t_{in} \leq 
t \leq  t_{in}+t_{w} \}. 
\end{equation}
The maximum value of $S_{max}\;[=S(t_{max})]$ 
depends on model parameters such
as the coupling strength ($K$), the noise intensity ($\beta$), 
the size of cluster ($N$), the coordination number ($Z$),
and the coupling randomness ($p$), 
as will be discussed 
in the following section.

\section{CALCULATED RESULTS}

\subsection{Regular couplings}

We have adopted same parameters of $\theta=0.5$,
$\alpha=0.5$, $\tau_s=10$, $A=0.10$, $t_{in}=100$
and $T_w=10$ as in \cite{Hasegawa03a}, and 
the $H$ function given by
\begin{equation}
H(x_{1j}-x_{1j})= x_{1j}-x_{1j}.
\end{equation}
DMA calculations have been made by solving Eqs. (22)-(35) with
the use of the fourth-order Runge-Kutta method with
the time step of 0.01.
We have performed also direct simulations by using  
the fourth-order Runge-Kutta method with the time step of 0.01.
Results of direct simulations are averages of 1000 trials for
$Z \le 20$ and those of 100 trials otherwise noticed.
All quantities are dimensionless.

First we discuss the case of regular couplings ($p=0.0$).
Figs. 2(a), 2(b), 2(c), and 2(d) 
show time courses of $\mu_1$, $\gamma_{1,1}$,
$\zeta_{1,1}$, and $\rho_{1,1}$, respectively,
with $\beta=0.005$, $K=0.02$,
$p=0.0$, $N=100$ and $Z=10$.
Results of DMA expressed by solid curves are in
good agreement with those of direct simulations
depicted by dashed curves.
Time courses of $\mu_1$, 
$\gamma_{1,1}$ $\zeta_{1,1}$, and $\rho_{1,1}$ 
shown in Fig. 2(a)-2(d)
are not so different from those for
sigmoid couplings reported in Figs. 3(a)-3(d) 
of \cite{Hasegawa04}.

Fig. 3(a)-3(c) show time courses of
$S(t)$ calculated by DMA (solid curves) 
and direct simulations (dashed curves) 
for $Z=10$, 50 and 99, whose magnitudes
are increased with increasing $Z$.
The maximum values of the
synchronization ratio in the DMA
are 0.0654, 0.386 and 0.569 for $Z=10$, 50 and 99, respectively,
which shows a larger synchrony for larger $Z$.
This is more 
clearly seen in Fig. 4(a) showing
$S_{max}$ as a function of $Z$.
Figure 4(b) shows the $Z$ dependences of
$\gamma_{1,1}$, $\zeta_{1,1}$ 
and $\rho_{1,1}$
at $t=t_{max}$
with $K=0.02$, $\beta=0.005$ and $N=100$:
filled and open marks express results of DMA and direct simulations,
respectively.
With increasing $Z$, $\gamma_{1,1}$ is significantly decreased 
while $\rho_{1,1}$ and $\zeta_{1,1}$
are almost constant.
This explains the larger synchrony $S_f$ for larger $Z$, 
shown in Figs. 4(a).
The difference between the $Z$ dependences of $\gamma_{1,1}$ and
$\rho_{1,1}$ is due to the fact that
$d \gamma_{1,1}/dt$ has a contribution 
from the second term of $2 K Z h_1(\zeta_{1,1}-\gamma_{1,1})$
in Eq. (24) while $d \rho_{1,1}/dt$
has no such contributions in Eq. (27).
Figure 4(b) shows that
$\zeta_{1,1}$ also depends on $Z$ 
because of the second term in Eq. (30).  
It is noted that because $t_{max}$ defined by Eq. (42)
depends on $Z$ in general, 
$\gamma_{1,1}$ at $t=t_{max}$ may show a weak $Z$
dependence, as shown in Fig. 4(b) where $t_{max}=107.16$, 106.72,
106.46 and 105.96 for $Z=10$, 20, 50 and 99, respectively.

\subsection{SW couplings}

Next we discuss the case of SW couplings by changing 
the coupling randomness $p$.
Figures 5(a), 5(b) and 5(c) show time courses of
$S(t)$ for $p=0.0$, 0.1 and 1.0, respectively, calculated 
by the DMA (solid curves) and direct simulations (dashed curves).
The maximum values of
the synchronization ratio $S_{max}$ in the DMA
are 0.0654, 0.0694, and 0.0749, for $p=0.0$, 0.1 and 1.0, 
respectively:
$S_{max}$ is slightly increased with increasing $p$.
This $p$ dependence of $S_{max}$ is more clearly seen in
Fig. 6(a) where $S_{max}$ is plotted against $p$
for $Z=10$.
Figure 6(b) shows the $p$ dependences of
$\gamma_{1,1}$, $\zeta_{1,1}$ 
and $\rho_{1,1}$ at $t=t_{max}$
with $K=0.02$, $\beta=0.005$, $N=100$ and $Z=10$:
filled and open marks express results of DMA and direct simulations,
respectively.
With increasing $p$, 
$\gamma_{1,1}$  
is slightly decreased while $\rho_{1,1}$ is not changed.
The origin of
the difference between the $p$ dependences of $\gamma_{1,1}$ and
$\rho_{1,1}$ is again due to the fact that
$d \gamma_{1,1}/dt$ has a contribution 
from the second term of $2 K Z h_1(\zeta_{1,1}-\gamma_{1,1})$
in Eq. (24) while $d \rho_{1,1}/dt$
has no such contributions in Eq. (27):
the $p$ dependence of $\gamma_{1,1}$ arises from 
$\zeta_{1,1}$ which depends on $p$ 
through network parameters of
$C$ and $R$ in Eq. (30), as shown in Fig. 6(b).

In Fig. 7, $S_{max}(p)$ normalized by its $p=0.0$ value
is plotted for various $Z$ with $N=100$, $K=0.02$ 
and $\beta=0.005$.
Values of $S_{max}(p=0.1)/S_{max}(p=0.0)$ in the DMA
are 1.061, 1.048, 1.0268 and 1.000
for $Z=10$, 20, 30, and 50, respectively:
an increase in $S_{max}$ is larger for smaller $Z$.

\section{CONCLUSION AND DISCUSSION}


Calculations in the preceding subsection show that
when the coupling randomness $p$ is introduced to
regular networks, the synchronization may be
slightly increased for diffusive couplings.
This is in strong contrast with the result
for sigmoid couplings in \cite{Hasegawa04}, which shows 
a decreased synchronization with increasing 
the coupling randomness.
The main origin of an increased synchronization
for diffusive couplings
may be their local negative feedback,
as will be discussed in the followings.
The diffusive coupling given by Eqs. (5) and (43)
may be rewritten as
\begin{eqnarray}
I_i^{(c)}(t)=K \sum_j c_{ij} (x_{1j}-x_{1i}) 
= K (\sum_j c_{ij} x_{1j} - k_i x_{1i})
\end{eqnarray}
where $k_i$ ($= \sum_j c_{ij}$) is heterogeneous. 
We may show that the heterogeneity in the
coordination number $k_i$ of Eq. (44) plays an important 
role in an increase of $S_{max}$ in SW networks.
If we replace $k_i$ by its average of
$Z \;(=<k_i>)$ in the feedback term of Eq. (44), it becomes 
\begin{equation}
I_i^{(c)}(t)
\sim K (\sum_j c_{ij} x_{1j} - Z x_{1i}).
\end{equation}
Filled and open circles in Fig. 8 denote  
$S_{max}$ calculated by using Eq. (44) 
with DMA and simulations, respectively,
for $N=100$, $Z=10$,
$\beta=0.005$ and $K=0.02$. 
Filled and open squares in Fig. 8 express  
$S_{max}$ calculated by using Eq. (45) 
with DMA and simulations, respectively,
for the same parameters as mentioned above.
Figure 8 clearly shows that
heterogeneous negative-feedback term 
($- K k_i x_{1i}$) in Eq. (44)
leads to a slightly increased synchronization whereas
the homogeneous one ($- K Z x_{1i}$) 
in Eq. (45) yields a decreased synchronization.

Equation (44) may alternatively be rewritten as
\begin{eqnarray}
I_i^{(c)}(t)&=& K \sum_j d_{ij} x_{1j},
\end{eqnarray}
with
\begin{eqnarray}
d_{ij}&=&c_{ij}-\delta_{ij} k_i.
\end{eqnarray}
It is noted that the new adjacent matrix $d_{ij}$ 
given by Eq. (47) satisfies the relation given by 
\begin{eqnarray}
\sum_j d_{ij} &=& 0.
\end{eqnarray}
Nishikawa {\it et al.} \cite{Nishikawa03}
have studied the stability of
synchronous states of coupled networks
in which adjacent (Laplacian) matrix is assumed to
satisfy the relation as given by Eq. (48). 
This implies that the coupling adopted in 
\cite{Nishikawa03} is related to a diffusive process.
From an analysis of the stability
of synchronous state by the Lyapunov index,
they have shown
that the synchronization becomes
more difficult in SW and SF networks with more heterogeneity.
Our calculation is expected not to be in contradict with
theirs because they examine the criteria for the stability
of synchronous oscillations while we have discussed
the degree of the synchronization for an applied signal.
It has been conventionally 
claimed that an increase in the synchronization
arises from the shorten average distance $L$ in SW networks
\cite{Fernandez00,Bucolo02,Hon02a,Hon02b,Buz04}.
However, equations of motions presented in
Eqs. (22)-(35) (and those in \cite{Hasegawa04}) 
do not include the term 
relevant to $L$ of SW networks. 

There is also the difference between effects
of heterogeneity for sigmoid and diffusive couplings.
For sigmoid couplings \cite{Hasegawa04}, 
the effect of the heterogeneity of
SW networks is included by a perturbation method 
with the term of 
$\delta c_{ij}\;[=c_{ij}(p)-c_{ij}(p=0)]$ through 
new correlations functions of $\phi_1$ and $\phi_2$
[see Eq. (37) in \cite{Hasegawa04}].
This has been made because the term of 
$< \delta x_{1i} \delta c_{ij}>$ appears in the process
of calculating equations of
motion, for example, of $d \gamma_{1,1}/d t$.
In contrast, for the diffusive couplings,
the counterpart term becomes 
$<\delta x_{1i} \delta x_{1j} \delta c_{ij}>$, which is
in the higher order than $< \delta x_{1i} \delta c_{ij}>$.
This shows that the effect of heterogeneity 
for diffusive couplings is weaker than that for
sigmoid couplings: for the diffusive couplings its effect
may be included by the $p$-dependent
$C$ and $R$ in the mean-field approximation,
while for sigmoid couplings it has to be taken into account
by the perturbation method.  
The stronger heterogeneity
for the sigmoid couplings
yields a decrease in the synchronization
when the heterogeneity is introduced.

To summarize, 
we have discussed the synchronization
in SW networks of spiking FN neurons
with diffusive couplings,
employing the semi-analytical DMA theory 
previously developed in \cite{Hasegawa04}.
A comparison of the results in this calculation with 
those for sigmoid couplings in \cite{Hasegawa04}
leads to the following conclusion.

\noindent
(1) When the average coordination number $Z$ is increased, 
the synchronization $S$ is increased 
both for sigmoid and diffusive couplings.
We should note, however, that
an increase in $S$ is mainly made by an increase in $\rho_{1,1}$ 
for sigmoid couplings [Fig. 4(a) in \cite{Hasegawa04}]
while it is accomplished by 
a decrease in $\gamma_{1,1}$ for diffusive couplings
[Fig. 4(b)].

\noindent
(2) When the coupling randomness
$p$ is increased, 
the synchronization $S$ is decreased
by an introduced heterogeneity for sigmoid couplings, 
whereas for diffusive couplings, $S$ may be slightly increased 
by their negative local feedback contribution
which compensates its decrease caused by their heterogeneity.

\noindent
It is noted that an increase in the synchronization 
for diffusive couplings
in the item (2) is not due to the shorten average distance 
in SW networks, against the conventional wisdom.
The item (2) is consistent with
the results in SW networks of the phase model
with the coupling term of $H(x-y)={\rm sin}(x-y)$, for which
an increase in the synchronization with increasing $p$
has been reported \cite{Hon02a,Hon02b}.
Items (1) and (2) imply that the synchronization
of SW networks depends not only on the geometry of SW networks
but also on details of couplings.
In the present paper, we have neglected 
the transmission time delay.
Because the average path length 
becomes shorter by added shortcuts
\cite{Wat98,Str01},
the response speed is expected to be improved 
in SW networks with time delays.
This is a great advantage of the SW networks
though the synchronization may be not necessarily improved.
Discussions in this paper and \cite{Hasegawa04} have been 
confined to SW neural networks
with symmetric (undirected) and unweighted couplings.
Recently it has been shown that the synchronization
in complex networks may be enhanced if their couplings
are undirected and weighted \cite{Mot05}.
It is interesting to apply our semianalytical 
approach to networks with
directed and weighted couplings, which are
realized in real complex networks.
This subject is left as our future study.

\section*{Acknowledgements}
This work is partly supported by
a Grant-in-Aid for Scientific Research from the Japanese 
Ministry of Education, Culture, Sports, Science and Technology.  

\newpage



\noindent
{\bf APPENDIX A: Derivation of Eqs. (22)-(35) }

Substituting Eq. (14) to Eqs. (3)-(6),
we get DEs for $\delta x_{1i}$ and $\delta x_{2i}$ of a neuron $i$,
given by (argument $t$ is suppressed)
\begin{eqnarray}
\frac{d \delta x_{1i}}{d t}
&=& f_1\:\delta x_{1i}+f_2 \:(\delta x_{1i}^2-\gamma_{1,1})
+ f_3 \: \delta x_{1i}^3 - c\: \delta x_{2i}  
+ \delta I_i^{(c)} +\xi_j, \\
\frac{d \delta x_{2j}}{d t}&=& b\:\delta x_{1j} - d \:\delta x_{2j},
\end{eqnarray}
with
\begin{eqnarray}
\delta I_i^{(c)}(t)
&=& K \:\sum_{j}  c_{ij} \:(h_1 \:[\delta x_{1j}(t)-\delta x_{1j}(t)]
+ h_3 \:[\delta x_{1j}(t)-\delta x_{1j}(t)]^3 ), 
\end{eqnarray}
where $f_{\ell}= (1/\ell !) F^{(\ell)}$ and $h_{\ell}= (1/\ell !) H^{(\ell)}$. 
DEs for the correlations are given by
\begin{eqnarray}
\frac{d \gamma_{\kappa,\lambda}}{d t}
&=& \left< \frac{1}{N} \sum_{i}
\left< \left[ \delta x_{\kappa i}\:
\left(\frac{d \delta x_{\lambda i}}{d t}\right)
+\left( \frac{d \delta x_{\kappa i}}{d t} \right)
\:\delta x_{\lambda i} \right] \right> 
\right>_c, \\
\frac{d \zeta_{\kappa,\lambda}}{d t}
&=& \left< \frac{1}{N Z} \sum_{i} \sum_{j} c_{ij} 
\left< \left[ \delta x_{\kappa i} \:
\left( \frac{d \delta x_{\lambda j}}{d t} \right)
+\left( \frac{d \delta x_{\kappa j}}{d t} \right)
\:\delta x_{\lambda i} \right] \right>
\right>_c,\\
\frac{d \rho_{\kappa,\lambda}}{d t}
&=& \left< \frac{1}{N^2} \sum_{i} \sum_{j} 
\left< \left[ \delta x_{\kappa i} \:
\left( \frac{d \delta x_{\lambda j}}{d t} \right)
+\left( \frac{d \delta x_{\kappa j}}{d t} \right) 
\:\delta x_{\lambda i} \right] \right>
\right>_c.
\end{eqnarray}
With the use of Eqs. (52)-(54), 
we may calculate DEs given by Eqs. (22)-(35).
For example, terms including $\delta I_i^{(e)}$ 
in $d \gamma_{1,1}/d t$,
$d \zeta_{1,1}/d t$ and $d \rho_{1,1}/d t$ become
\begin{eqnarray}
\left< \frac{2 }{N} \sum_i \left< \delta x_{1i} 
\delta I_i^{(c)} \right> \right>_c 
&=& \frac{2 K h_1}{N} \sum_i \sum_j 
c_{ij} \left< \left<\delta x_{1i} [\delta x_{1j}-\delta x_{1i} ]
\right> \right>_c, \\
&=& 2 K Z h_1 \: (\zeta_{1,1}-\gamma_{1,1}),\\
\left< \frac{2}{NZ} \sum_i \sum_j
c_{ij} \left< \delta x_{1i} \delta I_j^{(c)} \right> \right>_c
&=& \frac{2 Kh_1}{NZ} 
\sum_i \sum_j \sum_k
c_{ij}c_{jk} \left< \left< \delta x_{1i} 
[\delta x_k- \delta x_j] \right> \right>_c, \\
&=& \frac{2 K Z h_1}{N}[\gamma_{1,1}
+Z(C-R)\zeta_{1,1}+(ZR-ZC-1) \eta_{1,1}], \nonumber\\
&& \\
\left< \frac{2}{N^2} \sum_i \sum_j 
\left< \delta x_{1i} \delta I_j^{(c)} \right> \right>_c
&=&  \frac{2 K h_1}{N^2} \sum_i \sum_j \sum_k
c_{jk} \:\left< \left< \delta x_{1i} 
[\delta x_{1k}- \delta x_{1j}] \right> \right>_c, \\
&=& 0,
\end{eqnarray}
In evaluating Eqs. (55)-(60), 
we have employed the relation given by Eq. (13):
\begin{eqnarray}
\left< \left< \delta x_{\kappa i} \delta \:x_{\lambda j} \right> \right>_c
&=& \gamma_{\kappa,\lambda}\:\delta_{ij} 
+  \zeta_{\kappa,\lambda}\:c_{ij}\:  
+ \:\eta_{\kappa,\lambda}\:(1-\delta_{ij}-c_{ij}).
\end{eqnarray}

\newpage


\newpage

\begin{figure}
\caption{
The coupling randomness ($p$) dependence  
of (a) the clustering coefficient $C$
and (b) the coupling connectivity $R$  
of SW networks
for $Z=10$, 20 and 50 with $N=100$.
}
\label{fig1}
\end{figure}

\begin{figure}
\caption{
(color online).
Time courses of (a) $\mu_1$, (b) $\gamma_{1,1}$, (c) $\zeta_{1,1}$,
and (d) $\rho_{1,1}$
for $\beta=0.005$, $K=0.02$, $N=100$, $Z=10$ and $p=0.0$,
solid and dashed curves denoting results of DMA
and direct simulations, respectively.
At the bottom of (a), an input signal is plotted. 
Vertical scales of (b), (c) and (d) are multiplied
by factors of $10^{-4}$, $10^{-5}$ and $10^{-5}$,
respectively.
}
\label{fig2}
\end{figure}

\begin{figure}
\caption{
(color online).
Time courses of $S(t)$
for (a) $Z=10$, (b) $Z=50$ and (c) $Z=99$
calculated by DMA (solid curves) 
and direct simulations (dashed curves)
($\beta=0.005$, $K=0.02$, $N=100$ and $p=0.0$).
}
\label{fig3}
\end{figure}

\begin{figure}
\caption{
(color online).
The average coordination number ($Z$) dependence 
of (a) $S_{max}$ and
(b) $\gamma_{1,1}$ (circles), 
$\zeta_{1,1}$ (triangles) 
and $\rho_{1,1}$ (squares) at $t=t_{max}$
for $\beta=0.005$, $K=0.02$, $N=100$ and $p=0.0$:
filled and open marks denote results of DMA and
direct simulations, respectively.
}
\label{fig4}
\end{figure}

\begin{figure}
\caption{
(color online).
Time courses of $S(t)$
for (a) $p=0.0$, (b) 0.1 and (c) 1.0
calculated by DMA (solid curves) 
and direct simulations (dashed curves)
($\beta=0.005$, $K=0.02$ and $N=100$).
}
\label{fig5}
\end{figure}

\begin{figure}
\caption{
(color online).
The coupling randomness ($p$) dependence
of (a) $S_{max}$, and
(b) $\gamma_{1,1}$ (circles),
$\zeta_{1,1}$ (triangles)
and $\rho_{1,1}$ (squares) at $t=t_{max}$
for $\beta=0.005$, $K=0.02$, $N=100$ and $Z=10$:
filled and open marks denote results of DMA and
direct simulations, respectively. 
}
\label{fig6}
\end{figure}

\begin{figure}
\caption{
(color online).
The coupling randomness ($p$) dependence
of $S_{max}(p)/S_{max}(0)$
for $Z=10$, 20 and 50 and
with $\beta=0.005$, $K=0.02$, and $N=100$.
}
\label{fig7}
\end{figure}


\begin{figure}
\caption{
(color online).
The coupling randomness ($p$) dependence
of $S_{max}$ 
for the couplings $K \sum_j (c_{ij}-\delta_{ij} k_i) x_{1j}$ 
(circles) and $K \sum_j (c_{ij}-\delta_{ij} Z) x_{1j}$
(squares)
with $N=100$, $Z=10$, $\beta=0.005$, and $K=0.02$,
filled and open marks denoting results of DMA and
direct simulations, respectively
(see text).
}
\label{fig8}
\end{figure}

\end{document}